\documentclass[10pt]{article}
\usepackage[utf8]{inputenc}

\usepackage{listings}
\usepackage{xcolor}
\usepackage{amsfonts}
\usepackage{pifont}
\usepackage{physics}
\usepackage{amsmath}
\usepackage{geometry}
 \usepackage[ruled,vlined]{algorithm2e}

\definecolor{codegreen}{rgb}{0,0.6,0}
\definecolor{codegray}{rgb}{0.5,0.5,0.5}
\definecolor{codepurple}{rgb}{0.58,0,0.82}
\definecolor{backcolour}{rgb}{0.95,0.95,0.92}

\newtheorem{defnt}{Definition}[section]

\lstdefinestyle{mystyle}{
    backgroundcolor=\color{backcolour},   
    commentstyle=\color{codegreen},
    keywordstyle=\color{red},
    numberstyle=\tiny\color{codegray},
    stringstyle=\color{blue},
    basicstyle=\ttfamily\footnotesize,
    breakatwhitespace=false,         
    breaklines=true,                 
    captionpos=b,                    
    keepspaces=true,                 
    numbers=left,                    
    numbersep=5pt,                  
    showspaces=false,                
    showstringspaces=false,
    showtabs=false,                  
    tabsize=2
}

\newcommand{\Z}{\mathbb{Z}}
\newcommand{\Gr}{\mathbb{G}}
\newcommand{\cM}{\mathcal{M}}
\newcommand{\cR}{\mathcal{R}}
\newcommand{\cC}{\mathcal{C}}

\usepackage{graphicx}
\usepackage{tabularx}
\usepackage{authblk}
\usepackage{hyperref}
\usepackage{breakurl}
\usepackage[hang,flushmargin]{footmisc}

\newcommand\blfootnote[1]{%
  \begingroup
  \renewcommand\thefootnote{}\footnotetext{#1}%
  \addtocounter{footnote}{-1}%
  \endgroup
}

\newcommand{\protName}{FORT}
\title{FORT: Right-proving and Attribute-blinding \par Self-sovereign Authentication}
\date{}

\author[1,2]{Xavier Salleras}
\author[1]{Sergi Rovira}
\author[1,3]{Vanesa Daza}
\affil[1]{Department of Information and Communication Technologies, \par Universitat Pompeu Fabra, Barcelona, Spain}
\affil[2]{Dusk Network, Amsterdam, The Netherlands}
\affil[3]{CYBERCAT - Center for Cybersecurity Research of Catalonia \par \texttt{xavier@dusk.network, \{sergi.rovira, vanesa.daza\}@upf.edu}}

\begin{document}
\maketitle

\section*{Abstract}
\noindent 
Nowadays, there is a plethora of services that are provided and paid for online, like video streaming subscriptions, car or parking sharing, purchasing tickets for events, etc. Online services usually issue tokens directly related to the identities of their users after signing up into their platform, and the users need to authenticate using the same credentials each time they are willing to use the service. Likewise, when using in-person services like going to a concert, after paying for this service the user usually gets a ticket which proves that he/she has the right to use that service. In both scenarios, the main concerns are the centralization of the systems, and that they do not ensure customers' privacy. The involved Service Providers are Trusted Third Parties, authorities that offer services and handle private data about users. In this paper\blfootnote{This is the authors version of a work accepted and published in the special issue \textit{Advances in Blockchain Technology} of the journal \textit{Mathematics} (2022).}\makeatother, we design and implement FORT, a decentralized system that allows customers to prove their right to use specific services (either online or in-person) without revealing sensitive information. To achieve decentralization we propose a solution where all the data is handled by a Blockchain. We describe and uniquely identify users' rights using Non-Fungible Tokens (NFTs), and possession of these rights is demonstrated by using Zero-Knowledge Proofs, cryptographic primitives that allow us to guarantee customers' privacy. Furthermore, we provide benchmarks of FORT which show that our protocol is efficient enough to be used in devices with low computing resources, like smartphones or smartwatches, which are the kind of devices commonly used in our use case scenario.

\noindent
\\ \textbf{Keywords:} Zero-Knowledge Proofs; zk-SNARKs; Bulletproofs; Applied Cryptography; Self-sovereign; Internet of Things; Authentication; NFT.

\pagebreak 
\tableofcontents
\pagebreak 

\section{Introduction}
\label{sec:introduction}
Recently, smart cities have evolved with the inclusion of static internet-connected devices \cite{s18082507, 7395434, al2020intelligence} like pollution sensors, traffic lights, surveillance cameras, etc. Moreover, other mobile Internet of Things (IoT) \cite{9316004} devices, like autonomous cars, will be soon populating the cities. If we take into account all the computers, smartphones, smartwatches, etc., we observe how the density of devices is achieving high numbers. Technically speaking, handling this huge amount of connections will be possible thanks to 5G communications \cite{5gstandards}, which introduce \textit{network slicing} for splitting the network into many virtual and logical networks, providing specific features for different services. 

A high density of devices is also translated to more data shared over the network. A concerning fact is what happens with the data shared by users, especially when such data is sensitive or it can simply be used to profile users with no permission. Even when the usage of Internet technologies is increasing very fast, some security and privacy concerns \cite{ijaz2016smart, van2016privacy} still need to be addressed. In this paper, we address the problem of Trusted Third Parties (TTP), which are still required in many scenarios \cite{zhu2018asap}. For instance, GPS applications or autonomous driving are applications tracing our location and collecting much data about us. Even if the company behind says no personal data is collected, we can only trust them, with no possibility of detecting misbehavior. Other examples can be medical devices sharing sensitive information about patients through a trusted web server, or any other service that requires a TTP.

A natural solution to avoid the need for a TTP is the decentralization of its role. In this direction, several approaches \cite{8424682, di2018blockchain} raised in the last decade using Blockchain technologies to connect devices, skipping the need for a TTP in many use cases. In the same context, new digital services have appeared in the market, changing the way how users interact with them. Among many use cases, we can find car sharing, buying tickets for events, subscriptions to streaming services, etc. As centralization was a property of these applications that used to lead to control of all the network by some individuals, Blockchains \cite{blockchain2} started to change the way people interact with online services. The most common example, cryptocurrencies, has become a payment method without central authorities (i.e. banks) controlling the stream of the issued transactions and all the collateral information. Moreover, beyond being a payment solution, Blockchains like Ethereum \cite{ethereum} offer a way to execute programs \textit{on-chain}. Those programs, called \textit{smart contracts}, allow to issue a payment to a specific party as soon as this party proves that he/she meets some requirements specified in the contract. This same approach is used in many decentralized applications (DApps) \cite{dapps} nowadays, like paying the subscription to some service. 

\subsection{Motivation}

Decentralization implies that public data stored in the Blockchain can be accessed by anyone. This leads to some privacy concerns: as Blockchains publicly store all the network activity, user tracking or profiling becomes an issue to be addressed. In such regard, the problem gets worse when users of a Blockchain-based service need to interact with real-world services (i.e. proving to the staff of an event that you paid for the ticket), so if anyone learns your Blockchain identity, he learns all your history.

To solve the privacy concerns that arose on Blockchain applications, Zero-Knowledge Proofs (ZKP) started to be integrated within Blockchain projects like Zcash \cite{zcash}. ZKPs are cryptographic primitives allowing a user to prove to anyone else that some statement is true for some secret information, without leaking such information. In the Zcash example, these primitives allow users to issue transactions without leaking their identity nor the amount of money they are spending while proving that they are solvent.

Furthermore, Dusk Network \cite{dusk} is a Blockchain taking a step further and introducing a way to program \textit{private-by-design smart contracts}, programs whose execution and the parameters involved are kept private while being validated by the network. This leads to a toolset capable of building new privacy-preserving applications, solving many privacy and security concerns.

In such a scenario, the concept of \textit{self-sovereign authentication} \cite{Fedrecheski2020SelfSovereignIF} appeared: authentication systems where users can manage their identities in a fully transparent way, deciding which information they are willing to reveal to other parties. Some solutions like \verb!SANS! \cite{salleras2020sans} allow users to prove to Service Providers (SP) that they own a token that proves their right to use a specific service. Such a solution is suitable in many scenarios, but in some cases can have efficiency drawbacks since it relies on a ZKP construction called zk-SNARKs \cite{groth16}, which requires high computing power. This scheme is executable on Internet of Things (IoT) devices thanks to implementations like \verb!ZPiE! \cite{salleraszpie}, but taking a fair amount of time. This fact makes such a solution infeasible in use cases where IoT devices must prove several things in a short amount of time (i.e. willing to use a smartwatch to prove a right, having a door sensor with a cheap CPU verifying proofs, etc.).  Besides, this solution is still centralized, which means that if the SP disappears, the user no longer owns the right.

\subsection{Our Contributions}

In this paper, we introduce \verb!FORT!, a novel self-sovereign authentication protocol, combined with Blockchain technologies to provide a solution where users of a service acquire \textit{rights}, which are a set of different provable \textit{blinded attributes}. Such attributes are portions of personal information which have been blinded: they are invisible to the SP, and only the user can decide how much information about them has to be leaked. These attributes are represented by Non-Fungible Tokens (NFT) \cite{nfts} on the Blockchain, which can be granted \textit{on-chain} by entities providing services, the SPs, and verified \textit{off-chain}. For instance, a car willing to access a smart city would have to prove its right to do so, that is having two attributes: a certificate stating that the car has a low emissions level, and a fee payment receipt for entering the city. Once the right is represented in the Blockchain using an NFT, the car will be able to prove off-chain the possession of such a right, by using a cryptographic primitive called Zero-Knowledge Proof \cite{gmr85}. Such proof will state the possession of a valid NFT, without leaking the identifier of such NFT nor the identity of the car owner. Furthermore, our solution also skips third-party fees: for instance, in the scenario of buying tickets online for some event, in many cases, the ticket is issued by a third party who handles all the ticketing management, and who needs to be trusted. Furthermore, this party charges the users a service fee. Our solution relies on the Blockchain. Thus, the event organizer does not need to rely on third parties, and the user does not share his identity nor pay a service fee. 

Moreover, even when Zero-Knowledge Proofs require high computing resources, our solution can be deployed in IoT devices thanks to \verb!ZPiE! \cite{salleraszpie}, the library we used to implement a proof-of-concept of our solution. This allows users to use our protocol using devices with low computing resources such as smartwatches.

Our contribution relies on zk-SNARKs, but also on range proofs, another ZKP scheme where users prove that a value lies within a given range, without leaking such a value to other parties. In particular, we use the Bulletproofs \cite{bulletproofs} range proofs scheme. For that reason, our second contribution in this paper is the implementation of a Bulletproofs module for \verb!ZPiE!. Our implementation achieves excellent benchmarks, and using such a module, we implement our protocol and show its efficiency in IoT devices.

\subsection{Roadmap}

In Section \ref{sec:background}, we introduce the background needed to understand the paper. In Section \ref{sec:relatedwork}, we introduce the related work to our solution. In Section \ref{sec:zeroknowledge}, we introduce the cryptographic building blocks of our solution. In Section \ref{sec:oursolution}, we introduce \verb!FORT!, our solution. The implementation is detailed in Section \ref{sec:implementation}, along with its security analysis and several benchmarks. Section \ref{sec:futurework} discusses the future works that could be done to integrate our solution into more use cases. We conclude in Section \ref{sec:conclusions}.

\section{Background}
\label{sec:background}
In this section, we introduce the building blocks of our solution. We first introduce Blockchain technologies and their applications to IoT, and later we focus on the details and the specific use cases of smart contracts. We finally review Zero-Knowledge Proofs and how they are used to scale Blockchains by means of zk-Rollups.

\subsection{Blockchain}

A Blockchain \cite{blockchain} is a decentralized set of interconnected nodes, which share a unique and immutable set of data called ledger. Such a ledger is split into small portions of data called transactions, which are issued by different nodes in the network. In the scenario of cryptocurrencies like Bitcoin \cite{blockchain2}, these transactions are cryptographically validated by the network (i.e. a user sending some amount of bitcoins has enough funds to do so). Moreover, the whole network is ruled by a consensus agreed among all users of the network to keep the network safe (i.e. Proof-of-Work \cite{gervais2016security}, Proof-of-Stake \cite{bentov2014proof}, etc.). The decentralization properties of Blockchains and their security and privacy features led researchers to their integration with smart cities and IoT scenarios. It has been a hot research topic in recent years, for instance in surveys like the one provided in \cite{reyna2018blockchain}, where some of the challenges and opportunities in such regard are stated. Many use cases emerged in the last years, including renting, sharing, or selling specific assets, like cars or apartments. Other approaches are Blockchains applied to Wireless Sensor Networks \cite{cui2020hybrid} or e-health devices \cite{8167555}, among many others.

As such, huge amounts of data regarding IoT devices can be found in Blockchains, data that could be useful for improving smart cities traffic control, energy consumption, or pollution. Interesting approaches like the one proposed in \cite{7996641} allow for an easy IoT discovery in Blockchains. Regarding the security model, there are contributions like trust systems for IoT \cite{di2018blockchain}, where nodes of a Blockchain can be trusted upon checking their reputation, which changes depending on the behavior of the nodes in the network. 

\subsection{Smart contracts}

One of the most useful Blockchains in regards to our scenario is Ethereum \cite{ethereum}. It is a network whose purpose is not being a currency for making payments but a way to run distributed applications (DApps). DApps are possible thanks to smart contracts \cite{mavridou2018designing}, pieces of code executed on the Ethereum Virtual Machine (EVM) \cite{hildenbrandt2018kevm}. Such contracts and the EVM allow users, for instance, to be paid upon fulfilling some conditions. That is, for instance, distributed exchanges: applications where users buy or sell their cryptocurrencies to other users.

In order to execute transactions, Ethereum requires \textit{gas}. This is the amount of Ether (Ethereum's coin) per amount of bytes needed to run a transaction. Depending on how busy the Ethereum network is, the price of gas increases or decreases. This can make using Ethereum very expensive. To overcome such a problem, Zero-Knowledge-Rollups (zk-Rollups) have been proposed recently \cite{zkrollups}. They basically group several transactions into a single transaction of the main Ethereum Blockchain. Whereas the Ethereum network is called \textit{Layer 1}, the zk-Rollup is commonly called application of \textit{Layer 2}. zk-Rollups are possible thanks to Zero-Knowledge Proofs. Both Zero-Knowledge Proofs and zk-Rollups are introduced in Section \ref{subsec:zkp} and Section \ref{subsec:zkrollups} respectively.

Similarly, as Ethereum does, Dusk Network \cite{dusk} is a Layer 1 Blockchain that provides a virtual machine called \verb!Rusk! which enables the deployment and execution of smart contracts. However, they introduce the \textit{Confidential Security Contract Standard (XSC)}, which ensures the preservation of transactional confidentiality while simultaneously guaranteeing compliance through the use of Zero-Knowledge Proofs. This opens the door to a wide variety of use cases where privacy is a must, but accountability is required at the very same time.

\subsection{Zero-Knowledge Proofs}
\label{subsec:zkp}

\begin{figure}[ht]
  \centering
  \includegraphics[width=270pt]{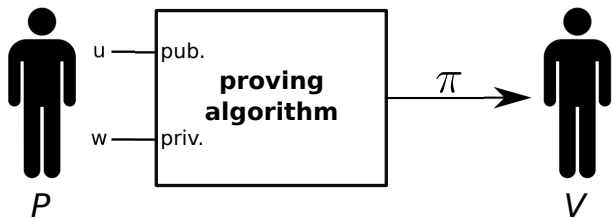}
  \caption{Zero-Knowledge Proof scenario.}
  \label{fig:zkp}
\end{figure}

A Zero-Knowledge Proof \cite{gmr85} is a cryptographic primitive which allows a prover $P$ to convince a verifier $V$ that a statement is true, without leaking any secret information. A statement is a set of elements known by both parties, defined as $u$, and the secret information only known by $P$ is called the witness $w$. $P$ wants to convince $V$ that he knows $w$, which makes a set of operations involving $u$, to hold. Such operations are defined by a \textit{circuit}, a graph composed of different wires and gates, which leads to a set of equations relating to the inputs and the outputs of these gates. Each of these equations is called \textit{constraint}. As depicted in Figure \ref{fig:zkp}, $P$ executes a proving algorithm using $u$ as the set of public inputs, and $w$ as the private inputs. This execution outputs a set of elements of an elliptic curve defined over a finite field, which we call the proof $\pi$. We send $\pi$ to $V$, who will use a verifying algorithm to verify that $u$ is true, for a given $w$ only known by $P$. Formally speaking, ZKPs must satisfy 3 properties:

\begin{itemize}
 \item \textbf{Completeness:} If the statement is true, $P$ must be able to convince $V$.
 \item \textbf{Soundness:} If the statement is false, $P$ must not be able to convince $V$ that the statement is true, except with negligible probability.
 \item \textbf{Zero-knowledge:} $V$ must not learn any information from the proof beyond the fact that the statement is true.
\end{itemize}

Even when first schemes required $P$ and $V$ to interact several times, Non-Interactive ZKPs (NIZKPs) \cite{Blum:1988:NZA:62212.62222} emerged, allowing $P$ to prove statements to $V$ by sending him a single message. However, the first schemes were expensive in terms of computing resources, and this made them not useful in real applications. More recently, zk-SNARKs appeared, which are Zero-Knowledge Succinct and Non-interactive ARguments of Knowledge \cite{cryptoeprint:2013:879}. This kind of proof is short and succinct: it can be verified in a few milliseconds, which makes it suitable for on-chain verification on Blockchains using smart contracts, being relatively cheaper in terms of gas consumption than other solutions. 

\begin{figure}[ht]
  \centering
  \includegraphics[width=390pt]{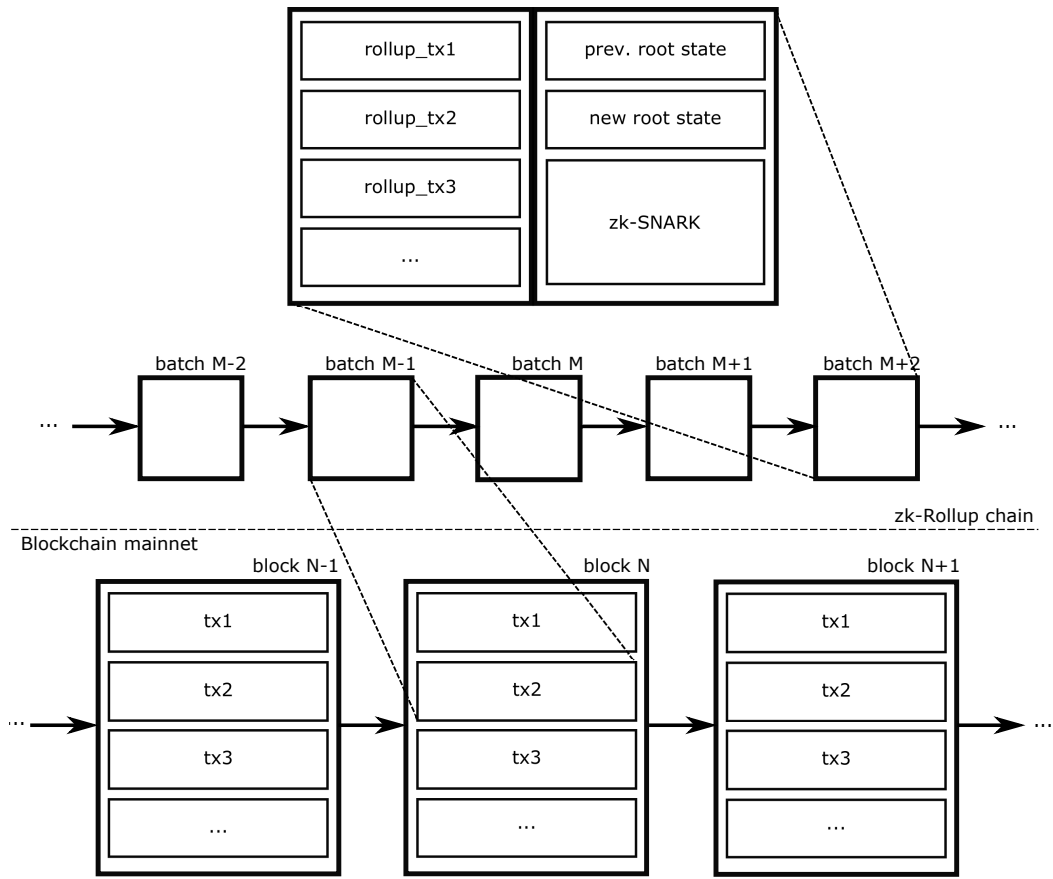}
  \caption{zk-Rollups overview.}
  \label{fig:zkrollup}
\end{figure}

\subsection{zk-Rollups}
\label{subsec:zkrollups}

zk-Rollups \cite{zkrollups}, as depicted in Figure \ref{fig:zkrollup}, create batches of several transactions in a Layer 2 scenario, and publish the whole batch into a single Layer 1 transaction. This saves a lot of gas that would be consumed if each transaction was executed directly on the main Blockchain. To do so, we have two actors, the \textit{transactors} willing to create a rollup transaction, and the \textit{relayers} computing the required operations to make the rollup work. In such regard, transactors send transactions to the relayers containing information about the sender, the receiver, the amount of tokens to be sent, etc. Such transactions also include a signature of the transaction. As stated previously, ZKPs require an elliptic curve, as proofs are sets of elements on such curves. For instance, the Barreto-Naehrig elliptic curve \cite{cryptoeprint:2005:133} called BN128, is the currently used curve for zk-SNARKs in Ethereum. The signature scheme used is EdDSA \cite{eddsa}, which also requires an additional elliptic curve where parameters are compatible with the zk-SNARKs elliptic curve (BN128). In this scenario, the Baby JubJub \cite{bjubjub} elliptic curve is used, for its compatibility with the parameters of BN128.

Once the relayer has received a bunch of transactions, he computes a Merkle tree of the previous accounts' state and the new state. Later, he computes a zk-SNARK which verifies all the signatures, and posts on the Blockchain a transaction containing this batch: the rollup transactions, the previous and new root states, and the zk-SNARK. This transaction is verified by a smart contract previously deployed on the Blockchain.

\section{Related work}
\label{sec:relatedwork}

Self-sovereign identity systems \cite{allen} have the premise of deploying protocols where users of different services can manage their identities in a secure, transparent, and private way. A general idea in this regard, and similar to our solution, was envisioned as a system where users can claim and prove possession of different rights associated with their identities, without compromising their privacy \cite{sovrin}. Furthermore, the combination of Self-sovereign identity systems with Zero-Knowledge Proofs has become a new research topic in the last years \cite{sov1}. In this regard, solutions like \verb!SANS! \cite{salleras2020sans} introduce a private authentication mechanism based on Zero-Knowledge Proofs. Using such tools, \verb!SANS! allows users to prove their rights to access several services, without the Service Provider (SP) knowing the identity of the users, while guaranteeing that the users are allowed to use the service (e.g. the users have paid a subscription fee).

There are some differences between \verb!SANS! and this work. In all cases, ownership of a given token can be proved (Proof of Ownership). Moreover, this work can prove that the token exists in the Blockchain (Proof of Transaction). Our solution is meant to have protection against malleability: we allow the Service Providers (SP) to be sure that a given token has been used only once. We also deploy attributes blinding, where our solution becomes completely self-sovereign: users reveal their data in a transparent and private way. Furthermore, the efficiency of our scheme is increased and its usage in IoT devices is totally feasible. 

Regarding privacy in online transactions, other research papers like \cite{9201400} explore an interesting way to provide a privacy-preserving authentication protocol by means of Physical Unclonable Functions (PUFs), providing a solid and efficient protocol.

Besides, ongoing research regarding how zk-SNARKs can contribute to Blockchains scalability is done in \cite{math9233016}, where research on distributed proofs generation making use of recursive zk-SNARKs is done.

On the other hand, combining IoT devices and NFTs is not an unexplored research area. Recent research \cite{secureiotnft} introduced a solution to manage IoT devices securely. They associate NFTs stored in Blockchains with IoT devices, to grant them a unique and indivisible identity.

Finally, to the best of our knowledge, there are no other solutions that provide a private-by-design and self-sovereign system to authenticate users, providing at the very same time a decentralized architecture, just like \verb!FORT! does.

\section{Cryptographic building blocks}
\label{sec:zeroknowledge}
In this section, we provide the necessary background on zk-SNARKs and Bulletproofs needed for the rest of the paper. We begin with a very high-level description of commitment schemes, and later move to an explanation of Bulletproofs and zk-SNARKs.

\subsection{Preliminaries}

We start by discussing a cryptographic primitive which is at the core of almost all modern cryptographic constructions, \textit{commitment schemes}. A commitment scheme allows us to select a secret value and commit to it, in the sense that the party performing the commitment cannot change that value for another in the future. The scheme gives the capability to reveal the value later on, but this is not a mandatory task. We are particularly interested in Non-interactive Commitment Schemes, defined as follows:

\begin{defnt}[Non-interactive Commitment Schemes]
A non-interative commitment scheme consists of a pair of probabilistic polynomial time algorithms (Setup, Commit). The setup algorithm $pp \leftarrow \text{Setup}(1^\lambda)$ generates public parameters $pp$ given the security parameter $\lambda$. Given the public parameters $pp$, the commitment algorithm Commit defines a function $\cM \times \cR \rightarrow \cC$ for message space $\cM$, randomness space $\cR$ and commitment space $\cC$. Given a message $x \in \cM$, the commitment algorithm samples $r \leftarrow \cR$ uniformly at random and computes $\text{Commit}(x;r) \in \cC.$
\end{defnt}

A useful commitment scheme for us is the Pedersen Commitment which we define as follows:

\begin{defnt}[Pedersen Commitment Scheme]
Let $\Gr$ be a group of order $p$ and set $\cM, \cR = \Z_p$ and $\cC = \Gr$. The Setup and Commit algorithms for Pedersen commitments are defined as follows:
\begin{itemize}
    \item Setup: Sample $g,h \leftarrow \Gr$ uniformly at random.
    \item $\text{Commit}(x;r)$: For a given $x \in \cM$, and a random value $r \leftarrow \cR$, we compute $g^x h^r \in \cC$.
\end{itemize}
\end{defnt}

\subsection{Bulletproofs}

Bulletproofs \cite{bulletproofs} are short non-interactive zero-knowledge arguments of knowledge that require no trusted setup. This means that the prover $P$ sends a single message to the verifier $V$, and this is enough to prove knowledge of the secret information. There is no need to rely on any prior information generated by a trusted party.

Bulletproofs were designed to enable efficient confidential transactions in cryptocurrencies, but they have found many other applications, such as shortening proofs of solvency or enabling confidential smart contracts \cite{cryptoeprint:2019:191}. The main technical feature of Bulletproofs is to prove that a committed value lies within a certain interval. For example, in the context of Blockchains, it is very useful to have an efficient protocol to prove that a secret value lies in the interval $[0,2^n-1]$ for some large value of $n \in \Z_{\geq 0}$. In the cryptographic community, this feature is called a \emph{range proof}. Range proofs allow us to prove that a secret value (previously committed to) lies within a certain range. They do not leak any information about the secret value but the fact that it lies within the desired range.

Let $\Gr$ be a cyclic group of prime order $p$ and let $\Z_p$ be the ring of integers modulo $p$. An \emph{inner-product argument} lets $P$ convince $V$ that he/she knows two vectors (bold font denotes a vector) $\vb{a},\vb{b} \in \Z_p^n$ such that $$C = \vb{g}^{\vb{a}} \vb{h}^{\vb{b}} \quad \text{and} \quad c = \langle \vb{a}, \vb{b} \rangle,$$
where $\vb{g},\vb{h}\in \Gr^n$ are independent generators, $c \in \Z_p$, and $C \in \Gr$. Now, let $c \in \Z_p$ and let $C \in \Gr$ be a Pedersen commitment to $c$ using randomness $r$. An inner-product range proof allows $P$ to convince $V$ that $c \in [0,2^{n}-1]$ by proving the relation
$$\{(g,h \in \Gr, C, n \ ; \ c, r \in \Z_p) \ : \ C = h^r g^c \wedge c \in [0, 2^{n} - 1] \}.$$

Now consider the case where $P$ needs to provide multiple range proofs at the same time. The idea of aggregated range proofs is to build a system that can provide a proof for multiple secret values and its efficiency is better than doing one proof for each of the secrets. Since the inner-product range proofs provided by Bulletproofs have logarithmic size, it is possible to build efficient aggregated logarithmic range proofs. That is, it is possible to efficiently prove the relation
$$\{ (g,h \in \Gr, \vb{C} \in \Gr^m \ ; \ \vb{c},\vb{r} \in \Z_p^m) \ : \ C_j = h^{r_j} g^{c_j} \wedge c_j \in [0,2^{n} - 1] \ \forall j \in [1,m] \},$$
where $m$ corresponds to the number of proofs. Bulletproofs can be computed in $O(n)$, and verified in linear time as well. The communication complexity (the size of the proofs) is $O(\log n)$.

\subsection{zk-SNARKs} 

One of the most used ZKP systems in practice are zk-SNARKs \cite{groth16}. This kind of proofs are short and succinct: they can be verified in only a few milliseconds. zk-SNARKs require a trusted setup that is used both by the prover and the verifier to generate and verify proofs. The set of parameters obtained during the setup phase is commonly called the Common Reference String (CRS). If an attacker was able to get the secret random values used to generate the CRS, it would be able to generate false proofs. For this reason, the initial setup is commonly made through a secure Multi-Party Computation (MPC) protocol \cite{cryptoeprint:2017:1050}, which generates the required parameters using a distributed computation protocol.

\begin{defnt}[zk-SNARKs]
A zero-knowledge succinct non-interactive argument of knowledge is a triple of algorithms (Setup, Prove, Verify) that works as follows:
\begin{itemize}
    \item $(pk, vk) \leftarrow \text{Setup}(1^\lambda, circuit)$: The setup algorithm  outputs a \emph{proving key} $pk$ and a \emph{verification key} $vk$ given the security parameter $\lambda$ and a circuit. Both keys (the CRS) are made public and can be used by the prover and the verifier to generate and verify proofs.
    
    \item $\pi \leftarrow \text{Prove}(pk, u, w)$: The proving algorithm produces a proof $\pi$ using the proving key $pk$ that attests that a statement $u$ and a witness $w$ are a correct solution to the set of equations derived from the circuit.
    
    \item $1/0 \leftarrow \text{Verify}(vk, u,\pi)$: The verification algorithm uses the verification key $vk$ to check whether $\pi$ is a correct proof for the public statement $u$.
\end{itemize}
\end{defnt}

zk-SNARKs can be computed in $O(n \log n)$. Both the verification and the communication complexity are $O(1)$.

\section{Our solution: \protName{}}
\label{sec:oursolution}
In this section, we introduce our solution with all details. We start with an overall description of our protocol, to later move to the specific details and its security analysis.

\subsection{Overall description}

\begin{figure}[ht]
  \centering
  \includegraphics[width=340pt]{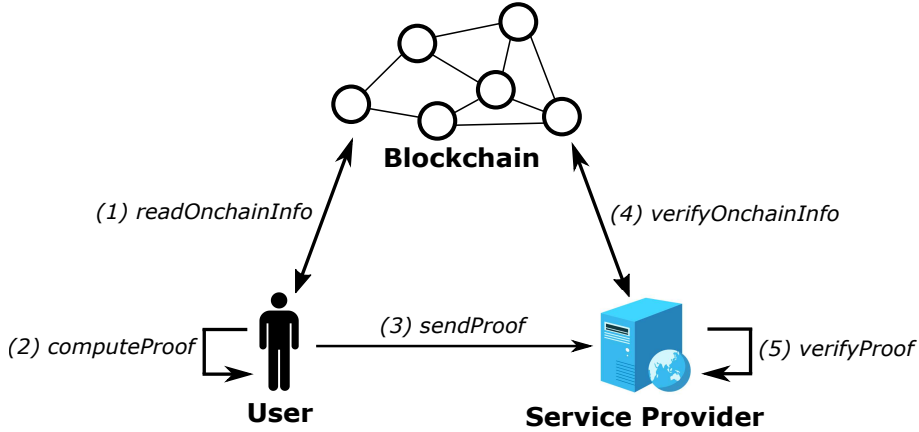}
  \caption{\protName{} protocol scenario overview.}
  \label{fig:scenario}
\end{figure}

Our solution is meant to be used in scenarios where users need to prove their right to use a service, for instance, accessing a house rented online. In such a scenario, we envision the usage of a certificate installed in the user's smartphone (or smartwatch, or any similar device) which can be validated by some sensor installed in the door of the house. Once validated, the SP might want a proof of meeting some requirements, linked with the previously validated certificate. We detail this workflow (as depicted in Figure \ref{fig:scenario}) in this section, as follows:

\begin{enumerate}
 \item \textbf{Read on-chain information:} the user acquires some attributes granted by third parties, which can be a SP to whom the user is buying a ticket or subscription, a governmental entity verifying your personal information, a bank providing a proof of solvency, etc. Such attributes are granted through an NFT stored in a Blockchain. The SP issues an NFT representing user's attributes. The SP mints this NFT on-chain, and later transfers it to the user's address. Now, the user can read these attributes from the Blockchain. 
 
 \item \textbf{Compute proof (the certificate):} the user acts as a prover, and computes a ZKP from the information collected from the NFT, as detailed in the circuit of Figure \ref{fig:circuit}, and installs this certificate in his device.
 
 \item \textbf{Send proof (read certificate):} The user tries to use the service by showing the certificate to the SP, who reads it.
 
 \item \textbf{Verify on-chain information:} the SP needs to partially read the Merkle tree of the Blockchain (as detailed in next section) to be able to verify (in the next step) that the attributes the user wants to prove are really on-chain (the NFT).
 
 \item \textbf{Verify proof (validate the certificate):} The SP verifies the ZKP, thus verifying the rights of the user. 
 
 \end{enumerate}
 
 After performing this protocol, the SP can ask the user for some information about the attributes, for instance, if they lay within a specific range. To do it, the user computes a bulletproof and sends it to SP, the verifier. Then, the SP verifies that the bulletproof sent by the user is correct, and knows for sure that the value is within a specific range.

Ideally, a desirable way to implement our protocol would be creating the NFTs directly using a Blockchain, and proving that we own them using the signature details involving the Blockchain transaction representing each NFT. However, this has some constraints regarding scalability and efficiency: first, the gas fees in the case of Ethereum can become very expensive, so using a transaction for a single right is far from being optimal. Second, the elliptic curve used to sign Ethereum transactions, the \textit{secp256k1}, is not pairing-friendly, so generating proofs proving ownership of the private key used to sign the transaction will not be efficient. To solve this problem, \verb!FORT! relies on zk-Rollups for scalability, and on EdDSA for proofs off-chain.

\subsection{Protocol details}

To be able to prove rights to access the service, the first thing a user needs to do is to receive an NFT stating some attributes about him. To do so, the user needs to contact the SP and provide him/her a proof of meeting some requirements. Then, SP executes Algorithm \ref{alg:createnft}: after validating the requirements, it issues an NFT representing the user's attributes. The SP mints this NFT on-chain, and later transfers it to the user's address. 

 \begin{algorithm}
\SetAlgoLined
	\textbf{Env:} vector of $x$ attributes: $attributes[x]$; user's address: $pk_{user}$; user's conditions: $cnd$ \par
	$nft \leftarrow create\_nft(attributes[x])$:\par
	\If{(verify\_conditions(cnd))}{
	  $nft.id = rand();$\par
	  $nft.attr = attributes[:];$\par
	  $nft.S \leftarrow sign_{sk_{SP}}(H(nft.id || nft.attr || pk_{user}));$\par
	  $mint\_nft(nft);$\par
	  $transfer\_nft(nft, pk_{user});$
  	}
 \caption{Create NFT}
 \label{alg:createnft}
\end{algorithm}

Upon receiving the NFT, the user is ready to anonymously prove possession of such an NFT. To do so, the user will follow the Algorithm \ref{alg:proveright}. Then, at some point the user will want to prove some rights, and to do so he/she will send / show the proof to the SP and it will execute Algorithm \ref{alg:verifycert}.

\begin{algorithm}
\SetAlgoLined
	\textbf{Env:} User and Service Provider (SP).
	\begin{enumerate}
	 \item The user reads the NFT transaction to be proved, which is published on the Blockchain, and the IDs of a set of NFT transactions in the batch $batch\_ids$, where $|batch\_ids| = 2^x$ and $x$ is agreed by consensus. 
	 \item The user requests access to the service offered by SP, and the SP sends a random value, a challenge $c$ to the user.
	 \item The user computes a proof $\pi$ using the above parameters as stated in the circuit depicted in Figure \ref{fig:circuit}.
	\end{enumerate}
 \caption{Create certificate}
 \label{alg:proveright}
\end{algorithm}

\begin{algorithm}
\SetAlgoLined
	\textbf{Env:} User and Service Provider (SP).
	\begin{enumerate}
	 \item The SP receives / reads $\pi$ from the user.
	 \item SP collects the ID of the transactions in the batch and computes the merkle tree $mtree$.
	 \item SP executes Algorithm \ref{alg:verifyright}, if it returns 1, SP grants the service.
	\end{enumerate}

 \caption{Verify certificate}
 \label{alg:verifycert}
\end{algorithm}

\begin{figure}[]
  \centering
  \includegraphics[width=470pt]{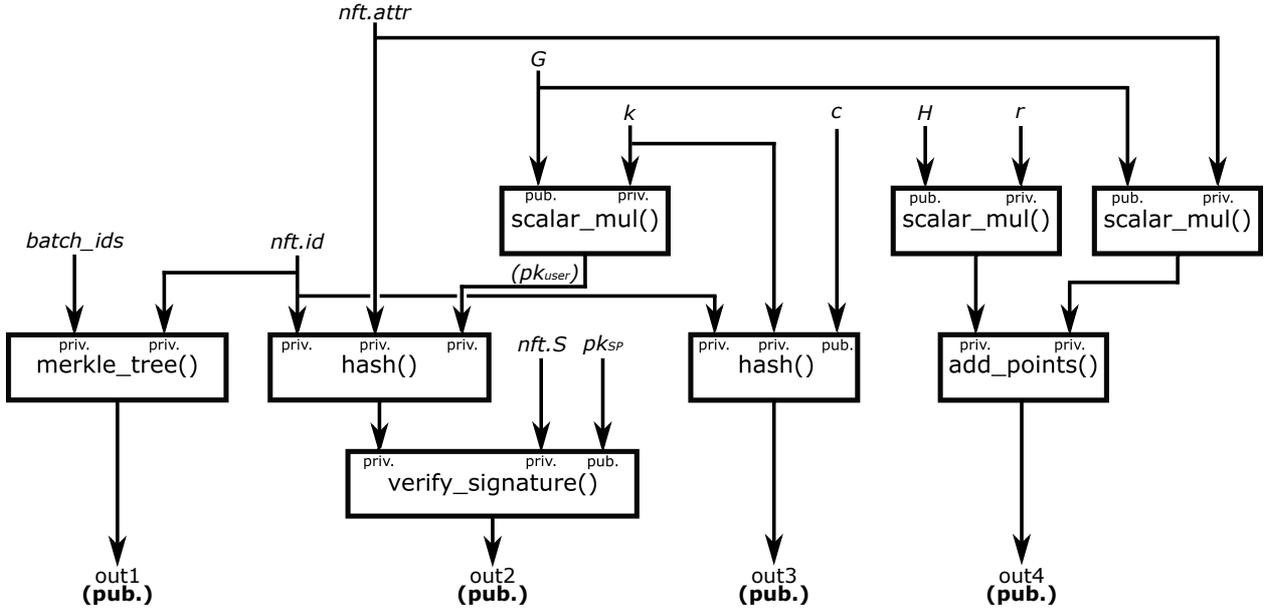}
  \caption{Circuit for our solution.}
  \label{fig:circuit}
\end{figure}

\begin{algorithm}
\SetAlgoLined
	\textbf{Env:} Zero-knowledge proof $\pi$.\par
	$1/0 \leftarrow verify\_right(\pi)$:\par
  	\eIf{(out1,out2,out3 $\leftarrow$ verify\_proof($\pi$)) and (out1 == mtree) and (out2 == 1) and !(is\_seen(out3))}{
  			return 1;
	}{
		return 0;
	}
 \caption{Verify right}
 \label{alg:verifyright}
\end{algorithm}

The proof used in Algorithm \ref{alg:proveright} uses the circuit depicted in Figure \ref{fig:circuit}. As can be seen, the circuit includes a verification of the signature $nft.S$, using the public key of the SP $pk_{SP}$. The inputs of the signature were the attributes $nft.attr$ along with the ID $nft.id$ and the public key of the user $pk_{user}$. The challenge $c$ is hashed along with $nft.id$ and the user's private key $k$. $nft.id$ is also hashed along with $batch\_ids$.

Finally, and before the SP grants the service, the prover might have to create a bulletproof to reveal some information about the attributes. This is done using $out4$, a Pedersen commitment. This process is detailed in Algorithm \ref{alg:createbulletproof}.

 \begin{algorithm}
\SetAlgoLined
	\textbf{Env:} secret key $k$; vector of $x$ attributes: $attributes[x]$;  vector of $x$ commitments: $C[x]$\par
	$\pi_b[x] \leftarrow create\_bulletproof(C[x], attributes[x])$:\par
	\For{i in x}{
	   $\pi_b[i] \leftarrow bulletproof(C[i], attributes[i])$
    }\par
 \caption{Create Bulletproof}
 \label{alg:createbulletproof}
\end{algorithm}

\subsection{Security analysis}

An important element to consider when analyzing the security of \verb!FORT! is the ZKP scheme to use. The main drawback of some ZKP constructions like zk-SNARKs, when used in scenarios like cryptocurrencies, is the need for a trusted setup. An untrusty setup could lead to huge losses of money if a malicious party gets the seed used to compute it, so it could create false transactions. This is not a problem in our solution: a different setup can be generated by each SP, as the proofs are verified off-chain by a single entity, the SP, and he is the main interested in not leaking the secret seed.

Moreover, the soundness property of each scheme relies on different security assumptions \cite{assumptions} (e.g. zk-SNARKs in \cite{cryptoeprint:2016:260} use a strong assumption, the q-Power Knowledge of Exponent ($q$-PKE) assumption). Furthermore, the security of these schemes relies on the security of elliptic curves, where breaking the security of the selected curve would lead to being able to generate false proofs. One of the most used curves in ZKPs is the BN128, which security level in practice is estimated to be 110-bits \cite{cryptoeprint:2016:1102}. Other curves like BLS12-381 \cite{zcash} estimate around 128-bits of security, with the drawback of heavier group operations. More recent research is introduced in \cite{cryptoeprint:2020:351}, where a new curve called BW6-761 is introduced. As stated by its authors, verification of proofs is at least five times faster than other state-of-the-art curves.

Regarding the circuit we have designed, our solution grants several privacy and authentication features:

 \begin{itemize}
 \item \textbf{Proof of Ownership:} the circuit used in \verb!FORT! verifies a signature $nft.S$ of an input $nft.id, nft.attr, pk_{user}$, using the public key of SP, $pk_{SP}$. Also, $pk_{user}$ is the output of the scalar multiplication $kG$, where $k$ is the user's private key. This ensures that the user owns the NFT, as only him can compute the public key using the private key, while keeping both values private, so SP cannot learn the identity of the user.
 
 \item \textbf{Proof of Transaction:} the circuit computes a Merkle tree of two private inputs: the $nft.id$ and the IDs of some other issued NFTs in the same batch $batch\_ids$. This ensures that the NFT the user is proving ownership of has been transacted in the Blockchain. SP can compute the Merkle tree $mtree$ itself, and check if it equals $out1$ as stated in Algorithm \ref{alg:verifyright}. 
 
 \item \textbf{Malleability protection:} the circuit computes the hash of $nft.id$, the private key $k$, and a challenge $c$. The format of this value could change in different scenarios. Taking the example of proving ownership of a ticket for an event, ideally, $c$ would be the date of such event. If $is\_seen(out3, previous) == 1$, it means that someone already entered the event with the same NFT. This is true because neither $nft.id$ nor $k$ can change, so $out3$ will always be the same for a given public input $c$. This prevents a user to use the same right multiple times, and to compute valid proofs for other users.
 
  \item \textbf{Attribute blinding:} the private information the user wants to share only when required, the attributes, are private inputs of the circuit. Such values are committed using a Pedersen commitment (i.e. $out4$, but as many as required can be included in the circuit), so the verifier learns these commitments, and the prover later uses a Bulletproof to prove knowledge of them, and to prove that they are within a specific range.
 
\end{itemize}

\verb!FORT!, as introduced in this section, can also be seen as framework to be modified to match the needs of every use case our solution could be deployed to. This means, selecting the proper ZKP scheme to be used, recompute the certificate each time instead of using Bulletproofs, select a different challenge $c$, etc.

\section{Implementation and benchmarks}
\label{sec:implementation}
In this section, we explain the capabilities and implementation details of the Bulletproofs module we developed, and later explain how we implemented our specific solution using our module.

\subsection{Bulletproofs module}

We implemented Bulletproofs as a module integrated into \verb!ZPiE!, a Zero-Knowledge Proofs library coded in C. The library uses GMP and MCL as dependencies: GMP is a pure C library used to handle big numbers and operations involving them, and MCL is a library written in C++, which offers a C wrapper for using it in pure C projects, used to do elliptic curve operations. The library also supports the elliptic curves BN128 and BLS12-381, which are thus also supported by our implementation. We implemented an API that allows us to generate aggregated range proofs using the Bulletproofs scheme above referred, and to verify them. The instructions on how to compile and use the library can be found in the README of the repository. The code can be used as explained in Listing 1.

\begin{lstlisting}[language=C, caption=Bulletproof generation example. Generation of 2 aggregated proofs of 64 bits.]
#include "../src/zpie.h"

int main()
{
    // init the bulletproofs module for 2 aggregated proofs of 64 bits
    bulletproof_init(64, 2);

    // set some values to prove knowledge of and compute the bulletproof 
    unsigned char *si[] = {"1234", "5678"};
    bulletproof_prove(si);

    // verify the bulletproof (../data/bulletproof.params)
    if(bulletproof_verify()) printf("Bulletproof verified.\n");
    else printf("Bulletproof cannot be verified.\n");
}
\end{lstlisting}

\begin{figure}[ht]
  \centering
  \includegraphics[width=280pt]{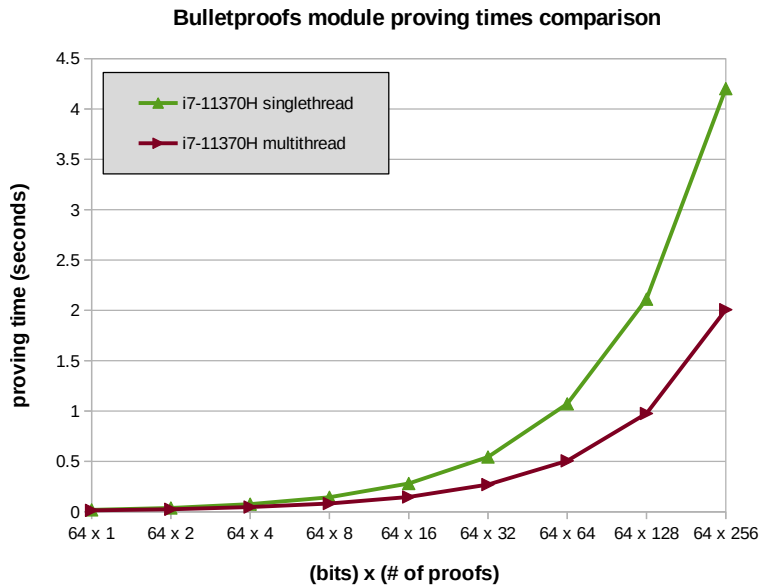}
  \caption{CPUs proving times of our solution.}
  \label{fig:charti7}
\end{figure}

We benchmarked our implementation as depicted in Figure \ref{fig:charti7}. Moreover, we improved the efficiency of our solution by using multi-threading in several parts of the prover and the verifier, where splitting the operations in different cores was possible. As we can see, we are benchmarking the time it takes by the prover, either in single-core (SC) or multi-core (MC), to compute the proofs. We performed the experiments for several amounts of aggregated proofs of 64 bits, using a 4-cores CPU, and the BN128.

\subsection{Solution deployment}

In this subsection, we detail the deployment of the three main parts of our protocol, \textit{generate rights}, \textit{generate the certificate}, and \textit{prove the attributes}.

\subsubsection{Generate rights}

The first step to use our solution is to generate the rights that our users will need to prove. To do so, a SP needs to provide a service and sell its subscription using an NFT minted to a smart contract-based Blockchain. For testing purposes, we used an Ethereum testnet where we created test NFTs using a reference implementation of ERC-721 (the Ethereum NFT standard)\footnote{https://github.com/nibbstack/erc721}. After deploying an NFT to the Blockchain, a user can buy it. Once done, he is ready to generate the certificate.

The computational costs for generating the NFTs are negligible, as no heavy cryptographic computations are involved in the process. Regarding the time it takes to be reflected on the Blockchain, it would depend on how crowded it is (typically it will take only a few minutes). On the other hand, one of the main concerns regarding this step when deploying it into the mainnet is the amount of gas required to execute the smart contract that mints the NFT. As discussed before, using zk-Rollups would be the best choice to reduce the cost when moving our solution to a production environment.

In Section \ref{sec:futurework} we discuss further work to be done regarding the deployment of our solution into a Blockchain network, considering how to boost even further the capabilities of our solution when using other Blockchains as the backbone of \verb!FORT!.

\subsubsection{Generate the certificate}

As explained previously, the prover precomputes the certificate, which is a zk-SNARK, required to use a specific service. The SP will verify the certificate and will be sure of the prover's right to use the service. We used \verb!circomlib!\footnote{https://github.com/iden3/circomlib} to estimate the number of constraints of the circuit used in our solution and thus, its efficiency. To create our circuit, we rely on four main functions:

\begin{itemize}
 \item \textit{scalar\_mul():} the circuit needs to multiply a number $k$ by a point on an elliptic curve $G$. To do this scalar multiplication using BN128, \verb!circomlib! uses \textbf{776 constraints}.
 \item \textit{hash():} the circuit needs to perform 2 fixed hashes, plus a variable number of hashes to compute a merkle tree. A fairly secure and efficient hash function is Poseidon \cite{cryptoeprint:2019:458}, which only uses \textbf{210 constraints} in \verb!circomlib!.
 \item \textit{verify\_signature():} we use the state-of-the-art signature scheme EdDSA \cite{eddsa} over BN128 provided in \verb!circomlib!, which uses \textbf{4018 constraints}.
 \item \textit{merkle\_tree():} the circuit needs to compute a merkle tree. Assuming that $|batch\_ids| = 256 = 2^8$, our solution will need to compute 8 Poseidon hashes. This sums up to \textbf{1680 constraints}. 
\end{itemize}

In total, our circuit can be implemented using \textbf{6894 constraints}. We coded a proof-of-concept using \verb!ZPiE!\footnote{https://github.com/xevisalle/zpie}, and executed the code using a laptop, a smartphone and a Raspberry Pi Zero. To demonstrate the scalability of our solution, we also executed the circuit using \verb!snarkjs!\footnote{https://github.com/iden3/snarkjs}, a JavaScript implementation of zk-SNARKs which can be executed in web browsers. This is perfect for scalability in web applications, with the performance drawback it involves, compared with binaries executed directly in the kernel. Table \ref{tab:results} shows the results.

\begin{table}[ht]
\centering
\caption{Performance results of  \protName{} in different devices using different implementations. All experiments use Groth'16 and BN128.}
\label{tab:results}
\vspace{0.1cm}
\begin{tabular}{c|c|c}
\hline
  \textbf{Device} & \textbf{Prover} & \textbf{Verifier}\\
  \hline
  Raspberry Pi Zero W (\verb!ZPiE!) & 79.058 s & 0.134 s\\
  \hline
  Snapdragon 732G (\verb!ZPiE!) & 0.830 s & 0.005 s\\
  \hline
  i7-11370H (\verb!ZPiE!) & 0.157 s & 0.000733 s\\
  \hline
  i7-11370H - Firefox (\verb!snarkjs!) & 0.694 s & 0.022 s\\
  \hline
\end{tabular}
\end{table}

As can be seen, either in high-end devices (a laptop CPU like i7-11370H) or in mobile CPUs (snapdragon 732G), the proofs used in our protocol can be computed in a fair small amount of time using \verb!ZPiE!. On the other hand, the time increases a lot when talking about extremely low-end CPUs like the one used in the Raspberry Pi Zero. Nevertheless, computing the proof in about a minute taking into account the single-core 700MHz CPU that it uses (aprox. 10\$), is a good result. Furthermore, an advantage of \verb!FORT! is that proofs can be precomputed much before to be used. Plus, even in worst-case scenarios, protocols like the one introduced in \cite{cryptoeprint:2018:691} would allow those devices to rely computations on other servers owned by the same user, using a secure channel. 

Regarding the verification of these proofs, as we stated previously, the verifier is succinct: all the proofs can be verified in just a few milliseconds, with no relation with the size of the circuit. As can be seen, \verb!ZPiE! outperforms here even in the Raspberry Pi, where it takes roughly 0.1 seconds to verify proofs.

Finally, we can see how either the prover and the verifier in \verb!snarkjs! are much slower than \verb!ZPiE! for the same CPU. However, such a result was expected, and taking into account the trade-off between performance and scalability, still it is a great result.

\subsubsection{Prove the attributes}

The SP, after verifying the certificate, might want to be sure that some of the attributes $nft.attr$ meet some additional requirements (e.g. being within a given range). For such purpose, we compute a Bulletproof from the Pedersen commitment described in the zk-SNARK circuit. We use the module introduced in the last section to achieve this outcome. In Listing 2 we show how to deploy our solution, where the prover proves knowledge of the Pedersen commitment, and that the secret lies within the range $[0,2^8 - 1]$.

\begin{lstlisting}[language=C, caption=Implementation of our solution.]
#include "../src/zpie.h"

int main()
{
    // we init the bulletproofs module, for a bulletproof of 8 bits
    bulletproof_init(8, 1);

    // we get the context (G, H, V[], gammas[])
    context ctx;
    bulletproof_get_context(&ctx);

    // we state that we will provide the random gamma and we assign it 
    // according to the one used in the certificate
    bulletproof_user_gammas(1);
    mclBnFr_setInt(&ctx.gammas[0], 1234); // r = 1234

    // we need to create a bulletproof for this commitment:
    // out4 = attr*G + r*H
    // we set the input attr = "250"
    unsigned char *si[] = {"250"};
    bulletproof_prove(si);

    // now P -> V: Bulletproof
    // V reads out4 from the certificate, and verifies the Bulletproof:
    if(bulletproof_verify()) printf("Bulletproof verified.\n");
    else printf("Bulletproof cannot be verified.\n");
}
\end{lstlisting}

The above code for proving knowledge of an 8-bit attribute takes only \textbf{0.3 seconds} on a Raspberry Pi Zero. This time increases as the size of the attributes does the same, but being a fair amount of time to be able to use our solution in IoT devices without problems. Executing the same approach using a zk-SNARK will require around 776 constraints, and the benchmark gives us \textbf{10.5 seconds}. As such, it is clear that Bulletproofs are a much better approach for this specific use case, where provers will be able to execute the protocol instantly using low-powered devices.

\section{Discussion on future works}
\label{sec:futurework}

We introduced a protocol that allows a user to get some rights to be used in different scenarios: the right to use a service (i.e. demonstrate to have some attributes like not being underage, having a salary above some threshold, etc.) or the right to access an event (i.e. demonstrate to have the attribute, in this case, the ticket for entering to an event, a performance, etc.). Our protocol works as-it-is in such scenarios. Needless to say, some changes shall be made if other constraints arise, or in other use cases. The usage of standard NFTs on Ethereum opens a wide range of features to implement. For instance, NFTs offer the feature of being transferred from one user to another, while charging a percentage of the selling price to the original creator of the token (e.g. the event planner). At the very same time, a SP organizing a performance could allow users to resell the tickets if they cannot attend, but prevent them to increase the price while preventing price speculations as well.

We have seen how \verb!FORT! could be easily deployed using Blockchains like Ethereum or Dusk. Regarding the latter, which at the time of writing is still under development, we have to take into account the private nature of the execution of their smart contracts. We envision how future work on a fully integrated solution within their network would lead to new privacy models, enhancing our protocol by even blinding the data we need to store on-chain.

\section{Conclusions}
\label{sec:conclusions}

In this paper, we introduced a protocol for proving ownership of a right for using a service or accessing an event, in a self-sovereign manner. Our protocol grants the chance to buy or request to be granted different attributes, which are grouped into rights, using a Blockchain. We can prove ownership of such rights using Zero-Knowledge Proofs, the main element of our \verb!FORT! protocol. After stating the details of \verb!FORT! and its security analysis, we performed several tests to show as using only 6894 constraints, it can be executed very efficiently in a wide variety of devices and environments: desktop, mobile, and web applications. As a reference, we have seen how we can compute a certificate with the users' rights in less than a second using a conventional smartphone. Later, the attributes of the certificate can be proved in just a few milliseconds. As future work, we discussed how our protocol could be modified to fit in other use cases, like ticket reselling or rights transferring, and we also discussed how integrating our solution into the Dusk Network Blockchain would lead to a higher level of privacy.

\phantomsection
\addcontentsline{toc}{section}{Acknowledgements}
\section*{Acknowledgements}
The authors are supported by Project RTI2018-102112-B-100 (AEI/FEDER, UE).

\phantomsection
\addcontentsline{toc}{section}{References}
\bibliographystyle{unsrt}
\bibliography{bibtex}

\end{document}